# The correlation timescale of the X-ray flux during the outbursts of soft X-ray transients

WU YuXiang[1,2], YU WenFei[1*] & LI TiPei[2]

[1] *Shanghai Astronomical Observatory, Shanghai 100012, China;*
[2] *Department of Engineering Physics & Center for Astrophysics, Tsinghua University, Beijing 100084, China*

Recent studies of black hole and neutron star low mass X-ray binaries (LMXBs) show a positive correlation between the X-ray flux at which the low/hard(LH)-to-high/soft(HS) state transition occurs and the peak flux of the following HS state. By analyzing the data from the All Sky Monitor (ASM) onboard the Rossi X-ray Timing Explorer (RXTE), we show that the HS state flux after the source reaches its HS flux peak still correlates with the transition flux during soft X-ray transient (SXT) outbursts. By studying large outbursts or flares of GX 339-4, Aql X-1 and 4U 1705-44, we have found that the correlation holds up to 250, 40, and 50 d after the LH-to-HS state transition, respectively. These time scales correspond to the viscous time scale in a standard accretion disk around a stellar mass black hole or a neutron star at a radius of $\sim 10^{4-5}\ R_g$, indicating that the mass accretion rates in the accretion flow either correlate over a large range of radii at a given time or correlate over a long period of time at a given radius. If the accretion geometry is a two-flow geometry composed of a sub-Keplerian inflow or outflow and a disk flow in the LH state, the disk flow with a radius up to $\sim 10^5\ R_g$ would have contributed to the nearly instantaneous non-thermal radiation directly or indirectly, and therefore affects the time when the state transition occurs.

**accretion disks, black hole physics, star: individual (Aquila X-1, GX 339-4, 4U 1705-44)**

Neutron star and black hole X-ray binaries show different spectral states in terms of spectral and timing characteristics [1, 2]. In the low/hard (LH) state, the spectrum is dominated by a hard, non-thermal power-law component which is probably due to the Comptonization of soft photons by hot optically thin plasma. The spectrum of the high/soft (HS) state is dominated by a thermal component probably from the inner region of an accretion disk [3]. Among these sources, a transient group of low mass X-ray binaries (LMXBs), namely the soft X-ray transients (SXTs), are important targets for the study of accretion physics because they usually go through a large range of mass accretion rates inferred from the observed X-ray flux range. During an outburst, an SXT normally goes through a state transition from the LH state to the HS state during the rising phase and a HS-to-LH state transition during the decay, though there are a few cases in which the LH-to-HS state transitions occur during outburst decays [4].

The observed X-ray spectra depend on the accretion process around a black hole or neutron star. This has been studied for many decades [5, 6]. In both early versions [7, 8] and later versions [9] of disk corona models, the corona was produced locally above or below the disk. More than ten years ago, Narayan et al. [10] proposed that the accretion flow in quiescent SXTs consists of two zones: an inner optically thin advection-dominated accretion flow (ADAF) extending from the central compact star to a transition radius, and an outer thin disk beyond. The model was further developed by Esin et al. [11] into a popular description of the spectral states. In all of these models, the disk corona or ADAF above the disk is assumed to be tied to the disk flow and corotate with the disk material, which however, is not necessary [12, 13].

A two-flow accretion geometry has been suggested which involves the presence of two dynamically independent components of the accretion flow at any radius, which propagate towards the compact star with different radial velocities [14–16]. Such a two-flow geometry might be formed when the accretion stream from the secondary splits into a disk part and an overshoot stream [17]. In this picture, a geometrically thick and optically thin sub-Keplerian halo flow, or a coronal flow, or an outflow, is above and below the Keplerian disk. The thermal component of the X-ray spectra is from the inner region of the Keplerian disk, and the non-thermal component is from the inner parts of the non-disk flow (e.g., from inverse Comptonization of soft photons in the hot plasma, or synchrotron radiation from an outflow). The same accretion rate variations are probably introduced to both flows at a large radius further out. These variations propagate on a sub-Keplerian time scale in the non-disk flow, and on the viscous time scale in the disk flow, respectively. The response of the thermal spectral component to these variations is expected to lag that of the non-thermal spectral component. This picture has been used to interpret the occurrence of the dim soft state [15, 18, 19] and the correlation between the LH-to-HS state transition flux and the HS state peak flux [4, 16, 20].

---

*Corresponding author (email: wenfei@shao.ac.cn)

One issue in accretion models is what parameters determine the transition between the spectral states. The transition from the LH state to the HS state is commonly thought to occur when the accretion rate increases, with the result of the inner disk moving inward [11]. However, the hysteresis phenomenon suggests that the accretion rate is not the only parameter which determines the state transition and that source history also matters [21–23]. There are other evidences, such as the occurrence of an LH-to-HS state transition while the flux is dropping in the LH state [4, 15], and the "parallel track" in the plot of the quasi-periodic oscillation frequency versus the X-ray flux for neutron star LMXBs [4, 24].

Recently, a nearly linear relation between the flux of the LH-to-HS state transition and the peak flux of the following HS state has been found in several sources in which multiple outbursts or flares with state transitions were covered by the RXTE observations [4, 16, 20], i.e., the higher the luminosity of the LH-to-HS transition is, the brighter the outburst will be, or vice versa. This has been confirmed by a study of the spectral transitions in the bright Galactic X-ray binaries observed with the X-ray monitoring observations of the RXTE/ASM and the Swift/BAT [25]. These studies suggest that the mass in the accretion disk appears to affect the flux at which a source transits from the LH state to the HS state. This can be explained by a two-flow geometry since the correlation requires a connection between the accretion flows producing different X-ray spectral components at different times. The accretion flow that powers the earlier LH state (i.e., produces the non-thermal radiation dominating the LH state) is somehow related to the accretion flow that powers the later HS state (i.e., produces the thermal radiation dominating the HS state). One possibility under the two-flow picture is that the non-thermal X-ray radiation in the LH state is influenced by the outer disk nearly instantaneously, while the thermal radiation from the disk flow only shows up in the X-rays when the disk flow propagates inward until it is close enough to the central star.

The time cost from the LH-to-HS state transition to the outburst flux peak during an outburst rise is similar among outbursts of the same source. According to previous studies [16, 20], the time costs are about 10, 30 and 100 d for Aql X-1, 4U 1705-44 and GX 339-4, respectively. These are the time scales where the transition flux and the outburst peak flux correlate. Under the two-flow picture, this correlation time scale reflects the range of the outer disk that contributes to the generation of the non-thermal X-ray radiation when the LH-to-HS transition occurs. One may speculate that the X-ray flux near the outburst peak correlates with the LH-to-HS transition flux and such a flux correlation would even hold in the early decay phase, i.e., the flux correlation holds longer than the time since the LH-to-HS state transition to the outburst peak.

Our goal in this work is to determine the time scale that the HS state flux remains correlated with the state transition flux after the source reaches its outburst peak. In order to fulfil this goal, we calculate Pearson's correlation coefficient between the ASM rate corresponding to the LH-to-HS state transition and the ASM rate of the following HS state at any given time. Considering the durations of the outbursts or flares are different, we intend to determine the longest correlation time scale for each source. It sets the outer disk range, which is physically allowed to be directly or indirectly involved in generating the non-thermal radiation in the LH state. We show that the correlation holds for as long as 40, 50, and 250 d for Aql X-1, 4U 1705-44 and GX 339-4, respectively, which corresponds to the viscous time scale in a standard disk up to a $\sim 10^{4-5}$ gravitational radius around the central compact star.

## 1 Data analysis

The RXTE/ASM data of the black hole LMXB GX 339-4 and the neutron star LMXBs Aql X-1, 4U 1705-44 in the period from January 1996 to March 2007 were used in our study. The ASM light curves were first averaged every 3 days. The LH-to-HS state transition is determined by the ratio between the hard X-ray flux and soft X-ray flux. During the rise of the outburst, the hardness ratio will exhibit a sharp decrease at a certain point, which corresponds to the LH-to-HS transition. Conventionally, the soft X-ray refers to the radiation below 10 keV and hard X-ray above 10 keV. Thus, it is best to use the PCA (2–9 keV) data and HEXTE (15–259 keV) data to respectively represent soft and hard X-ray. However, RXTE pointed observations cannot always cover the LH-to-HS transition. Whenever PCA and HEXTE observations were not available, we calculated the count rate ratios between the three ASM channels, i.e., 1.3–3.0, 3.0–5.0, and 5.0–12.2 keV, which were used to determine the LH-to-HS transition time. The energy region of ASM does not cover the non-thermal radiation very well, and the detection area is small. Therefore, the uncertainty of the state transition decided by ASM will be larger than that by PCA and HEXTE.

We calculated the Pearson's correlation coefficient between the ASM rate of the HS state at any given time $\Delta t$ after the LH-to-HS transition and the ASM rate at which the state transition occurred for the outbursts or flares of a given source. It is known that in the LH state just before the LH-to-HS state transition, the power-law spectral index is in a narrow range for the same source [26]. So, the 2–12 keV ASM rate at the start of the LH-to-HS state transition represents the X-ray transition flux. Therefore, the correlation coefficients we calculated indicate whether there is a correlation between the state transition flux and the flux of the HS state at a certain time after the transition.

The relation between the correlation coefficient and the time separation $\Delta t$ is therefore obtained for each source. We chose $\Delta t$ from 0 with a step size of 3 d. When we calculated the correlation coefficient for a certain $\Delta t$, only the outbursts whose HS states extended beyond $\Delta t$ are taken into the account. When the number of the outburst samples is smaller than three, the calculation ends. We got the coefficient for the longest $\Delta t$ when the samples decreased to three. Consequently, the longest time scale that we can study depends on the three longest outbursts for each source.

In order to derive statistical uncertainties of the correlation coefficients, we simulated 100 light curves corresponding to each ASM light curve by redistributing the ASM count rates according to their values and uncertainties. We calculated the correlation coefficient for each of the simulated ASM light curves in relation to $\Delta t$, then averaged the results from the 100 simulated light curves, and derived the corresponding uncertainties of the averages.

## 2 Results

The determination of the LH-to-HS state transition time for each outburst or flare is a crucial step in our analysis. The transition time can be determined within a day or so with the ratio between the HEXTE (15–250 keV) count rate and the PCA (2–9 keV) count rate [20]. Transition times in the three outbursts of GX 339-4 are known from pointed observations [20], which occurred around MJD 50813, 52397 and 53233. Yu & Dolence [4] have also identified observations corresponding to the LH-to-HS state transition in four outbursts of Aql X-1 with the RXTE pointed observations. The third outburst among the four, which is special because it was dim and with an LH-to-HS state transition occurring during a flux decline, is excluded from our analysis. The LH-to-HS state transition times of the other three outbursts are MJD 51316, 51820 and 53162. We also determined that the HS states of the outbursts ended on MJD 51164, 52721, and 53536 for GX 339-4, and MJD 51364, 51865, and 53177 for Aql X-1, respectively. The ASM rates after these end dates were not used in our calculation.

As mentioned before, RXTE pointed observations were not able to cover the LH-to-HS state transition in every outburst. If we want to study as many outbursts as we can, we have to decide state transitions from ASM hardness ratios. For GX 339-4, four major outbursts occurred in the past decade. We found that the state transition times of the first three outbursts according to ASM hardness ratios are consistent with those based on pointed observations [20] (Figure 1). The fourth outburst, which occurred in 2007, is included in our study and the state transition was determined from ASM hardness ratios. Its rising and declining time scales are much shorter than the previous three. The state transition times of the three outbursts in Aql X-1 have been derived from pointed observations [4]. For the other outbursts of Aql X-1 and all the flares of 4U 1705-44, we estimated the LH-to-HS transition times from the ASM hardness ratios. However, in a few cases, the uncertainties of ASM colors are too large to determine the exact transition times. We then selected the ASM hardness ratio which corresponds to an ASM rate nearest to 4 c/s as the transition time. This brings uncertainties into the transition times of only a few days. The times when the HS-to-LH state transition occurred were estimated as well. The HS-to-LH state transition usually occurs at a very low flux where the uncertainties in the ASM hardness ratios are large. In consideration that the HS-to-LH state transition generally occurred at a relatively constant flux level, we chose the end time of the HS state as the time when the ASM could not detect the source for the first time during the outburst decay at $1\sigma$, i.e., the uncertainty in the ASM rate is comparable to or larger than the ASM rate itself.

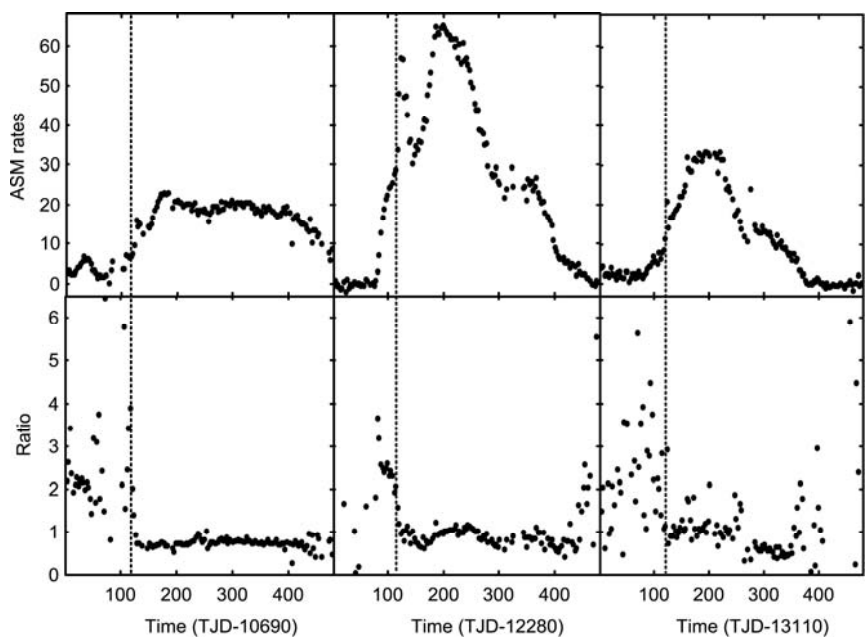

**Figure 1**  Upper row: ASM light curves of the three outbursts of GX 339-4 studied in Yu et al. [20]; lower row: the corresponding ASM color of (5.0–12.1 keV)/ (1.3–5.0 keV). Dashed lines indicate the LH-to-HS transition time for each outburst in this study.

In the end, we found 8 outbursts in Aql X-1 (upper panel in Figure 2, including the three outbursts whose state transition times were studied with pointed observations in Yu & Dolence [4]) and 12 flares in 4U 1705-44 (lower panel in Figure 2), excluding a few flares in 4U 1705-44 because the ASM data during these flares were too sparse to identify the LH-to-HS state transitions. We then calculated the correlation coefficients as introduced before. It is worth noting that the uncertainties in the state transition times determined from the ASM data could be from a few days (e.g., for Aql X-1) to a week (e.g., for 4U 1705-44).

We plot the correlation coefficient versus $\Delta t$ for GX 339-4, Aql X-1, and 4U 1705-44 in Figure 3. The upper left panel shows the result for the first three outbursts of GX 339-4. The coefficients are close to unity when $\Delta t$ is less than 100 d, then drop to zero when $\Delta t$ exceeds 200 d. The upper right panel shows the result after taking the 2007 outburst into account. Because its duration (about 100 d) is less than half of the durations of the previous three, the dramatic drop of the correlation coefficients at $\Delta t$ less than 100 d can be seen. At $\Delta t$ larger than the duration of the fourth outburst, only the first three outbursts contribute. So, the correlation coefficient curve is the same as on the left. For Aql X-1, the analysis of 8 outbursts based on the ASM data as well as pointed observation data extend $\Delta t$ to above 40 d. The results show that the flux correlation does not hold when $\Delta t$ is more than 40 d. The correlation coefficient versus $\Delta t$ of 4U 1705-44 presented in the lower right panel shows a similar behavior. We also calculate the correlation between the HS peak flux and the HS flux, which is not significantly better (even worse for 4U 1705-44) than the correlation between the LH-to-HS transition flux and the HS flux.

Besides the HS and LH state, GX 339-4 also experienced the very high state (VHS) or intermediate state (IS) during the initial sharp rise of the 2002 outburst. The IS hardness ratio is between those of the LH state and the HS state. The LH-to-HS state transition of the 2002 outburst, based on the hardness ratio, falls right into the IS defined in other literatures [27, 28]. We did not discriminate IS in our study, because there is no evidence that IS delays the HS state peak. Actually, the duration of the intermediate state is relatively short, and the time separation between the LH-to-HS transition and the HS state peak was similar no matter if the outburst experienced IS or not. However, it did reduce the correlation coefficient. In the upper left panel in Figure 3, the correlation coefficient is apparently lower than unity at small Δt (< 30 d), and rises to unity after IS. Except for this single case, the evolution behaviors of outbursts studied in this work can be well described by the simply-defined LH and HS state, which reflects the relative proportion of non-thermal radiation and disk thermal radiation. The full description of spectral states given in McClintock & Remillard [2] is not relevant to this paper. Homan et al. [28] divided the 2002 outburst of GX 339-4 into the classic LH, IS and HS state, and pointed out that the HS state includes the observations that would be classified as a steep power law or thermal dominant state defined by McClintock & Remillard [2].

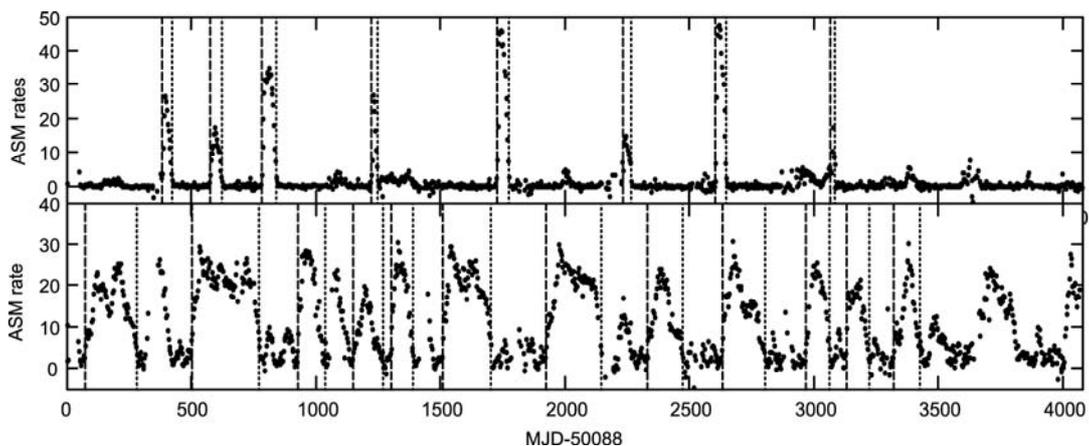

**Figure 2** The ASM (2–12 keV) light curves of Aql X-1 (upper) and 4U 1705-44 (lower) in the past 10 years. Eight (including the three outbursts of Aql X-1 studied in Yu & Dolence [4] with pointed observations) and twelve relatively bright outbursts are selected respectively for each source, with the dashed lines marking the LH-to-HS state transition times and dotted lines marking the end times of the HS states of the outbursts. The bright outbursts of 4U 1705-44 occurred around the time (MJD-50088) of 350, 1100, and 3700 d are excluded because the ASM did not cover the rising phases when the LH-to-HS state transitions should have occurred.

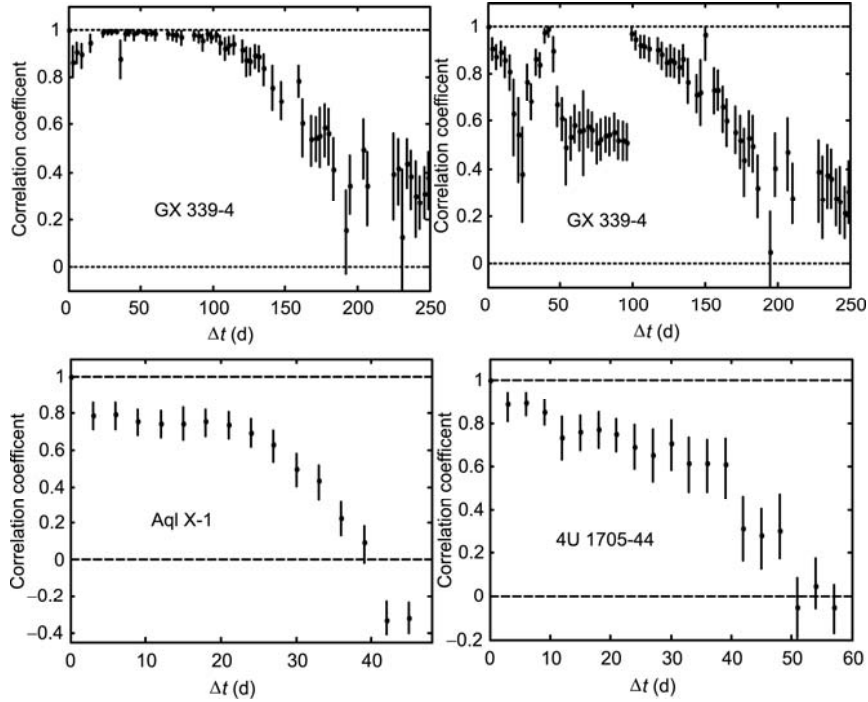

**Figure 3** The relation between the correlation coefficient and time separation Δt between the LH-to-HS state transition and the following HS state for GX 339-4 (upper), Aql X-1 (lower left), and 4U 1705-44 (lower right). The relations are derived from three outbursts of GX 339-4 (upper left, see Figure 1, studied by Yu et al.[20]), four outbursts of GX 339-4 (upper right, the 3 outbursts on the left as well as a recent one, see text), eight outbursts of Aql X-1 (see upper panel in Figure 2), and twelve flares of 4U 1705-44 (see lower panel in Figure 2), respectively.

A positive correlation coefficient at a certain time separation Δt indicates that the higher the flux of the LH-to-HS state transition is, the higher the flux of the HS state at time Δt after the state transition is. From this plot, we can also estimate the longest time scale on which the flux correlation holds, i.e. the corresponding Δt where the correlation coefficient remains positive. They are about 250, 40 and 50 d for GX 339-4, Aql X-1 and 4U 1705-44, respectively. These time scales are significantly longer than the time separation between the LH-to-HS transition and the HS state peak inferred from previous studies, i.e., approximately 100, 10 and 30 d, respectively.

## 3 Discussions and conclusion

We have calculated the correlation coefficient between the flux at which the LH-to-HS state transition which occurred and the HS state flux at a certain time after the transition in the black hole transient GX 339-4 and two neutron star LMXBs Aql X-1 and 4U 1705-44. We have found that the flux correlation holds even after the source has passed the flux peak in the HS state, which were up to 250, 40 and 50 d for GX 339-4, Aql X-1 and 4U 1705-44, respectively. These time scales are longer than those between the state transitions and the HS state flux peaks by a factor of about two or more.

It is known that in the HS state, the disk flow dominates [2, 3]. We are able to estimate the time scale that a perturbation in the mass accretion rate at a certain location in the outer disk propagates to the center. This is the viscous or radial drift time scale at a certain radius in a standard α disk. The viscous time scale is expressed as [29]

$$t_{\rm visc} = 3\times10^5 \alpha^{-4/5} \dot{M}_{16}^{-3/10} M_1^{1/4} R_{10}^{5/4}\, s.$$

The radius in the disk corresponding to a certain viscous time scale is then

$$\frac{R}{R_g} \approx 4\times 10^5 \left(\frac{t_{\rm visc}}{100\,{\rm d}}\right)^{4/5} \left(\frac{\alpha}{0.2}\right)^{16/25} \left(\frac{\dot{M}}{0.2\dot{M}_{\rm Edd}}\right)^{6/25} \left(\frac{M}{M_\odot}\right)^{-24/25},$$

in which $R_g = 2GM/c^2$ is the gravitational radius. In order to make an order-of-magnitude estimate, we take both $\alpha$ and $\dot{M}$? as 0.2, because in the HS state, $\alpha$ should be large [29] and should be above the mass accretion rate threshold around 0.01–0.1 [11]. The masses of the compact objects in the three sources are unknown. Assuming $M\sim 10 M_\odot$ for GX 339-4 (GX 339-4 has a mass function of $5.8 M_\odot$), $M\sim M_\odot$ for Aql X-1 and 4U 1705-44, and taking $t_{\rm visc}$ as the maximal correlation time scale in our study, the disk radii corresponding to the maximal correlation time scales of the three sources are $9.0\times 10^4\, R_g$ or $2.7\times 10^{11}$ cm for GX 449-4, $1.4\times 10^5\, R_g$ or $4.1\times 10^{10}$ cm for Aql X-1, and $1.6\times 10^5\, R_g$ or $4.7\times 10^{10}$ cm for 4U 1705-44, respectively. They are all in the order of $10^5\, R_g$. So, the longest time scale on which the correlation holds suggests that the instantaneous mass accretion rate in the disk flow at a radius up to $10^5\, R_g$ correlates with the instantaneous mass accretion rate in the LH state when the LH-to-HS transition starts.

This is an important piece of information that helps determine the accretion geometry [14–16, 20, 24] in the LH state. A characteristic of the proposed two-flow geometry is that at a certain radius, there exist two dynamically different accretion flows, i.e., a Keplerian disk flow and a sub-Keplerian non-disk flow (which could be a corona flow or an outflow in recent works), and the disk flow is sandwiched by the non-disk flow which is above and below the disk. The two flows have different response times to the variation of mass accretion rate at the outer edge of the accretion system, ranging from that of free-fall to longer time scales expected from different suggestions. Finally, the two flows are related in mass accretion rate, either established at the outer edge of the disk flow or through a coupling of the two flows not limited only to the outer disk edge.

Our results and the results from previous studies of the state transitions in LMXB [4, 16, 20] can be interpreted under the two-related-flow geometry. First, the fact that the hard X-ray peak precedes the soft X-ray peak supports a two-flow geometry, because the non-thermal radiation powered by the non-disk flow responds quickly to the variation of accretion rate on a nearly free-fall time scale [15, 19] or several times shorter than the viscous time scale in the disk flow [4, 16], while the thermal radiation powered by the disk flow responds on a viscous time scale. Second, the flux correlation is difficult to explain if the two flows are absolutely independent. The linear correlation between the flux of the LH-to-HS state transition and that of the soft X-ray peak flux, and the correlation between the transition flux and the HS flux after the soft X-ray peak, indicate that the disk flow producing the soft X-rays in the HS state and the non-disk flow producing the hard X-rays at much earlier times are related in terms of accretion rate.

We have shown that the observed longest time scale on which the flux correlation holds corresponds to the viscous time scale in a standard disk at a radius of $10^5\, R_g$, indicating the maximal radius where the two flows are initially related. If the two flows are supplied with a proportional amount of matter at the outer disk edge as discussed in Smith et al. [15], the radius of $10^5\, R_g$ corresponds to the radius of the outer disk edge where the two flows split. In general under the two-flow geometry, since the non-disk flow produces the non-thermal spectral component which dominates in the LH state, we can infer that the outer disk with a radius as large as $10^5\, R_g$ might affect the non-thermal radiation instantaneously and play a role during the LH-to-HS transition. This size is comparable to or larger than the disk size estimated in other literatures [15, 18, 30], but in the same order of the tidal radius in these binary systems which is likely the maximal radius allowed for in an accretion disk based on theory [29]. GX 339-4 has an orbital period of 42 h. Assuming the primary mass is $10 M_\odot$ and the mass ratio of 0.1, the tidal radius is then about $3\times 10^{11}$ cm. Aql X-1 has an orbital period of 19 h. If we take the primary mass as 1 $M_\odot$ and the mass ratio of 0.5, the tidal radius is about $1\times 10^{11}$ cm. The estimates are of the same order of magnitude as the radius derived from the correlation time scale. It is worth noting that if the duration of the HS state corresponds to the viscous time scale in the accretion disk, a shorter outburst should have a smaller disk.

*This work was supported by the National Natural Science Foundation of China (Grant Nos. 10533020, 10773023, and 10833002), the One Hundred Talents project of the Chinese Academy of Sciences, the Shanghai Pujiang Program (Grant No. 08PJ14111), the National Basic Research Program of China (Grant No. 2009CB824800), and the Starting Funds at the Shanghai Astronomical Observatory. The study has made use of the data obtained through the High Energy Astrophysics Science Archive Research Center Online Service, provided by the NASA/Goddard Space Flight Center.*